\documentstyle[12pt,amsbsy]{article}

\newcommand{\mrm}[1]{\mathrm{#1}}
\newmathalphabet*{\mbf}{cmr}{b}{n}
\newmathalphabet*{\mii}{cmr}{m}{it}
\newmathalphabet*{\mtt}{cmtt}{m}{n}

\newcommand{\alphaem}{\alpha_{\mrm{em}}}

\newcommand{\pT}{p_{\perp}}
\newcommand{\pTmin}{p_{\perp\mrm{min}}}

\newcommand{\pom}{\mrm{I}\!\mrm{P}}
\newcommand{\reg}{\mrm{I}\!\mrm{R}}

\renewcommand{\b}{{\mathrm b}}
\renewcommand{\c}{{\mathrm c}}
\renewcommand{\d}{{\mathrm d}}
\newcommand{\e}{{\mathrm e}}
\newcommand{\f}{{\mathrm f}}
\newcommand{\g}{{\mathrm g}}

\newcommand{\p}{{\mathrm p}}
\newcommand{\q}{{\mathrm q}}
\newcommand{\s}{{\mathrm s}}

\renewcommand{\u}{{\mathrm u}}

\newcommand{\K}{{\mathrm K}}

\newcommand{\pbar}{\overline{\mathrm p}}
\newcommand{\qbar}{\overline{\mathrm q}}

\newcommand{\Jpsi}{\mrm{J}/\psi}
\newcommand{\ee}{\e^+\e^-}

\newenvironment{Itemize}{\begin{list}{$\bullet$}%
{\setlength{\topsep}{0.2mm}\setlength{\partopsep}{0.2mm}%
\setlength{\itemsep}{0.2mm}\setlength{\parsep}{0.2mm}}}%
{\end{list}}
\newcounter{enumct}
\newenvironment{Enumerate}{\begin{list}{\arabic{enumct}.}%
{\usecounter{enumct}\setlength{\topsep}{0.2mm}%
\setlength{\partopsep}{0.2mm}\setlength{\itemsep}{0.2mm}%
\setlength{\parsep}{0.2mm}}}{\end{list}}

\newlength{\captivewidth}
\newcommand{\captive}[1]{\rule{5mm}{0mm}%
\begin{minipage}{\captivewidth}%
\caption[small]{#1}\end{minipage}}

\begin{document}

\sloppy

\pagestyle{empty}

\begin{flushright}
CERN-TH.7193/94
\end{flushright}

\vspace{\fill}

\begin{center}
{\LARGE\bf $\boldsymbol{\gamma}\boldsymbol{\gamma}$ and
$\boldsymbol{\gamma}$p Events at
High Energies$^*$} \\[10mm]
{\Large Gerhard A. Schuler and Torbj\"orn Sj\"ostrand} \\[3mm]
{\large Theory Division, CERN} \\[1mm]
{\large CH-1211 Geneva 23}\\[1mm]
{\large Switzerland}\\
\end{center}

\vspace{\fill}

\begin{center}
\bf{Abstract}
\end{center}
\vspace{-0.5\baselineskip}
\noindent
A real photon has a complicated nature, whereby it may remain
unresolved or fluctuate into a vector meson or a perturbative
$\q\qbar$ pair. Based on this picture, we previously
presented a model for $\gamma\p$ events that is based on the
presence of three main event classes: direct, VMD and anomalous
\cite{gammap}.
In $\gamma\gamma$ events, a natural generalization gives
three-by-three combinations of the nature of the two incoming
photons, and thus six distinct event classes. The properties
of these classes are constrained by the choices already made, in
the $\gamma\p$ model, of cut-off procedures and other aspects.
It is therefore possible to predict the energy-dependence of
the cross section for each of the six components separately.
The total cross section thus obtained is in good agreement with
data, and also gives support to the idea that a simple factorized
ansatz with a pomeron and a reggeon term can be a good approximation.
Event properties undergo a logical evolution
from $\p\p$ to $\gamma\p$ to $\gamma\gamma$ events, with larger
charged multiplicity, more transverse energy flow and a higher
jet rate in the latter process.

\vspace{\fill}
\noindent
\rule{6cm}{0.4mm}

\vspace{3mm} \noindent
{\Large {\bf$^*$}}Presented by T. Sj\"ostrand at the Workshop on
Two-Photon Physics from DA$\Phi$NE to LEP200 and Beyond,
Paris, France, 2--4 February 1994

\vspace{\fill}

\noindent
CERN-TH.7193/94 \\
March 1994

\clearpage

\pagestyle{plain}
\setcounter{page}{1}

\section{Introduction}

There are many reasons for being interested in $\gamma\gamma$ physics.
The process $\ee \to \ee\gamma\gamma \to \ee X$ will be a main one
at LEP 2 and future linear $\ee$ colliders. Therefore, $\gamma\gamma$
events are always going to give a non-negligible
background to whatever other physics one is interested in. However, more
importantly, the collision between two photons provides the richest
spectrum of (leading-order) processes that is available for any choice
of two incoming elementary particles. For instance, since the photon
has a hadronic component, all of hadronic physics is contained as a
subset of the possibilities. A correct description of the components
of the total $\gamma\gamma$ cross section is therefore the ultimate
challenge of `minimum-bias' physics.

The study of $\gamma\gamma$ physics has a long history, and it is not
our intention here to give a complete list of references.
Many topics have been covered by the
contributions to this and other workshops \cite{workshop}.
For the approach we are going to take,
one important line of work is the subdivision of photon interactions
by the nature of the photon \cite{PWZ}. Minijet phenomenology has
attracted much attention in recent years \cite{minijet}.

However, none of these approaches attempts to give a complete
description of $\gamma\gamma$ cross sections and event properties,
but only concentrate on specific topics. Here we will try to be
more ambitious, and really provide all the necessary aspects in one
single framework.
The starting point is our model for $\gamma\p$ physics \cite{gammap},
which can be generalized in an (almost) minimal fashion. The results
presented here are preliminary, in the sense that a number of further
cross-checks are in progress \cite{ourgaga}.

One main area is still left out of our description: in all that
follows, both incoming photons are assumed to be on the mass shell.
Further issues need to be addressed when either photon or both of them
are virtual. For reasons of clarity, we also restrict ourselves to
discussing what happens in the collision between two photons of given
momenta. The addition of photon flux factors will complicate the
picture, but not add anything fundamentally new.

\section{Event Classes}

To first approximation, the photon is a point-like particle. However,
quantum mechanically, it may fluctuate into a (charged)
fermion--antifermion pair. The fluctuations
$\gamma \leftrightarrow \q\qbar$ are of special interest to us,
since such fluctuations can interact strongly and therefore turn
out to be responsible for the major part of the $\gamma\p$ and
$\gamma\gamma$ total cross sections, as we shall see. On the other
hand, the fluctuations into a lepton pair are uninteresting, since
such states do not undergo strong interactions to leading order, and
therefore contribute negligibly to total hadronic cross sections. The
leptonic fluctuations are perturbatively calculable, with an
infrared cut-off provided by the lepton mass itself. Not so for
quark pairs, where low-virtuality fluctuations enter a domain of
non-perturbative QCD physics. It is therefore customary to split
the spectrum of fluctuations into a low-virtuality and a high-virtuality
part. The former part can be approximated by a sum over low-mass
vector-meson states, customarily (but not necessarily) restricted
to the lowest-lying vector multiplet. Phenomenologically, this
Vector Meson Dominance (VMD) ansatz turns out to be very successful in
describing a host of data. The high-virtuality part, on the other hand,
should be in a perturbatively calculable domain.

In total, the photon wave function can then be written as
\begin{equation}
|\gamma\rangle = c_{\mrm{bare}} |\gamma_{\mrm{bare}}\rangle +
\sum_{V = \rho^0, \omega, \phi, \Jpsi} c_V |V\rangle +
\sum_{\q = \u, \d, \s, \c, \b} c_{\q} |\q\qbar\rangle +
\sum_{\ell = \e, \mu, \tau} c_{\ell} |\ell^+\ell^-\rangle
\label{gammawavefunction}
\end{equation}
(neglecting the small contribution from $\Upsilon$). In general, the
coefficients $c_i$ depend on the scale $\mu$ used to probe the photon.
Thus $c_{\ell}^2 \approx (\alphaem/2\pi)(2/3) \ln(\mu^2/m_{\ell}^2)$.
Introducing a cut-off parameter $p_0$ to separate the low- and
high-virtuality parts of the $\q\qbar$ fluctuations, one similarly
obtains $c_{\q}^2 \approx (\alphaem/2\pi) 2e_{\q}^2 \ln(\mu^2/p_0^2)$.
The VMD part corresponds to the range of $\q\qbar$ fluctuations below
$p_0$ and is thus $\mu$-independent (assuming $\mu > p_0$).
In conventional notation
$c_V^2 = 4\pi\alphaem/f_V^2$, with $f_V^2/4\pi$ determined from data
to be 2.20 for $\rho^0$, 23.6 for $\omega$, 18.4 for $\phi$ and
11.5 for $\Jpsi$ \cite{Baur}. Finally, $c_{\mrm{bare}}$ is given by
unitarity: $c_{\mrm{bare}}^2 \equiv Z_3 = 1 - \sum c_V^2 -
\sum c_{\q}^2 - \sum c_{\ell}^2$. In practice, $c_{\mrm{bare}}$ is
always close to unity. Usually the probing scale $\mu$ is taken to be
the transverse momentum of a $2 \to 2$ parton-level process. Our fitted
value $p_0 \approx 0.5$ GeV (see below) then sets the minimum transverse
momentum of a perturbative branching $\gamma \to \q\qbar$.

\begin{figure}[t]
\captive%
{Contributions to hard $\gamma\p$ interactions: a) VMD,
b) direct, and c)~anomalous. Only the basic graphs are illustrated;
additional partonic activity is allowed in all three processes.
The presence of spectator jets has been indicated by dashed lines,
while full lines show partons that (may) give rise to
high-$\pT$ jets.
\label{Fig1}}
\end{figure}

The subdivision of the above photon wave function corresponds to the
existence of three main event classes in $\gamma\p$ events,
cf. Fig.~\ref{Fig1}:
\begin{Enumerate}
\item The VMD processes, where the photon turns into a vector meson
before the interaction, and therefore all processes
allowed in hadronic physics may occur. This includes elastic and
diffractive scattering as well as low-$\pT$ and high-$\pT$
non-diffractive events.
\item The direct processes, where a bare photon interacts with a
parton from the proton.
\item The anomalous processes, where the photon perturbatively branches
into a $\q\qbar$ pair, and one of these (or a daughter parton thereof)
interacts with a parton from the proton.
\end{Enumerate}
All three processes are of $O(\alphaem)$. However, in the direct
contribution the photon structure function is of $O(1)$ and the
hard scattering matrix elements of $O(\alphaem)$, while the opposite
holds for the VMD and the anomalous processes.
As we already noted, the $\ell^+\ell^-$ fluctuations are not
interesting, and there is thus no class associated with them.

The above subdivision is not unique, or even the conventional one. More
common is to lump the jet production processes of VMD and anomalous
into a class called resolved photons. The remaining `soft-VMD' class
is then
defined as not having any jet production at all, but only consisting
of low-$\pT$ events. We find such a subdivision
counterproductive, since it is then not possible to think of the
VMD class as being a scaled-down version (by a factor $c_V^2$) of
ordinary hadronic processes --- remember that normal hadronic
collisions {\em do} contain jets part of the time.

In a complete framework, there would be no sharp borders between the
three above classes, but rather fairly smooth transition regions that
interpolate between the extreme behaviours. However, at our current
level of understanding, we do not know how to do this, and therefore
push our ignorance into parameters such as the $p_0$ scale and the
$f_V^2/4\pi$ couplings. From a practical point of view, the sharp
borders on the parton level are smeared out by parton showers and
hadronization. Any Monte Carlo event sample intended to catch a border
region would actually consist of a mixture of the three
extreme scenarios, and therefore indeed be intermediate. Also, remember
that our separation is
applied to leading-order processes,
with higher-order effects included only in the leading-logarithmic
approximation.
An additional scheme dependence would arise for truly
higher-order matrix elements.

The difference between the three classes is easily seen in terms
of the beam jet structure. The incoming proton always gives a beam jet
containing the partons of the proton that did not interact. On the
photon side, the direct processes do not give a beam jet at all, since
all the energy of the photon is involved in the hard interaction. The
VMD ones (leaving aside the elastic and diffractive subprocesses for the
moment) give a beam remnant just like the proton, with a `primordial
$k_{\perp}$' smearing of typically up to half a GeV. The anomalous
processes give a beam remnant produced by the $\gamma \to \q\qbar$
branching, with a transverse momentum going from $p_0$ upwards.
Thus the transition from VMD to anomalous should be rather smooth.

\begin{figure}[t]
\captive%
{Contributions to hard $\gamma\gamma$ interactions: a) VMD$\times$VMD,
b)~VMD$\times$direct, c) VMD$\times$anomalous,
d) direct$\times$direct, e) direct$\times$anomalous,
and f)~anomalous$\times$anomalous.
Notation as in Fig.~\protect\ref{Fig1}.
\label{Fig2}}
\end{figure}

A generalization of the above picture to $\gamma\gamma$ events is
obtained by noting that each of the two incoming photons is
described by a wave function of the type given in
eq.~(\ref{gammawavefunction}). In total, there are therefore
three times three
event classes. By symmetry, the `off-diagonal' combinations appear
pairwise, so the number of distinct classes is only six.
These are, cf. Fig.~\ref{Fig2}:
\begin{Enumerate}
\item VMD$\times$VMD: both photons turn into hadrons, and the processes
are therefore the same as allowed in hadron--hadron collisions.
\item VMD$\times$direct: a bare photon interacts with the partons of the
VMD photon.
\item VMD$\times$anomalous: the anomalous photon perturbatively
branches into a $\q\qbar$ pair, and one of these (or a daughter parton
thereof) interacts with a parton from the VMD photon.
\item Direct$\times$direct: the two photons directly give a quark pair,
$\gamma\gamma \to \q\qbar$. Also lepton pair production is allowed,
$\gamma\gamma \to \ell^+\ell^-$, but will not be considered by us.
\item Direct$\times$anomalous: the anomalous photon perturbatively
branches into a $\q\qbar$ pair, and one of these (or a daughter parton
thereof) directly interacts with the other photon.
\item Anomalous$\times$anomalous: both photons perturbatively branch
into $\q\qbar$ pairs, and subsequently one parton from each photon
undergoes a hard interaction.
\end{Enumerate}
The first three classes above are pretty much the same as the three
classes allowed in $\gamma\p$ events, since the interactions of a VMD
photon and those of a proton are about the same.

The main parton-level processes that occur in the above classes are:
\begin{Itemize}
\item The `direct' processes $\gamma\gamma \to \q\qbar$ only occur
in class 4.
\item The `1-resolved' processes $\gamma\q \to \q\g$ and
$\gamma\g \to \q\qbar$ occur in classes 2 and 5.
\item The `2-resolved' processes $\q\q' \to \q\q'$ (where $\q'$
may also represent an antiquark), $\q\qbar \to \q'\qbar'$,
$\q\qbar \to \g\g$, $\q\g \to \q\g$, $\g\g \to \q\qbar$ and
$\g\g \to \g\g$ occur in classes 1, 3 and 6.
\item Elastic, diffractive and low-$\pT$ events occur in class 1.
\end{Itemize}
The notation direct, 1-resolved and 2-resolved is the conventional
subdivision of $\gamma\gamma$ interactions. The rest
is then called `soft-VMD'. As for the $\gamma\p$
case, our subdivision is an attempt to be more precise and internally
consistent than the conventional classes allow. One aspect is that
we really want to have a VMD$\times$VMD class that is nothing but
a scaled-down copy of the $\rho^0\rho^0$ and other vector-meson
processes, with a consistent transition between low-$\pT$ and
high-$\pT$ events (see below). Another aspect is that, in a
complete description, the VMD and anomalous parts of the photon give
rise to different beam remnant structures, as discussed above, even
when the hard subprocess itself may be the same.

A third aspect is that our subdivision provides further constraints;
these, at least in principle, make the model more predictive. In
particular, the parton distributions of the photon are constrained
by the ansatz in eq.~(\ref{gammawavefunction}) to be given by
\begin{equation}
f_a^{\gamma}(x,\mu^2) = f_a^{\gamma,\mrm{dir}}(x,\mu^2)
+ f_a^{\gamma,\mrm{VMD}}(x,\mu^2)
+ f_a^{\gamma,\mrm{anom}}(x,\mu^2;p_0^2) ~.
\label{gammaPDF}
\end{equation}
Here
\begin{equation}
f_a^{\gamma,\mrm{dir}}(x,\mu^2) = Z_3 \, \delta_{a\gamma} \,
\delta (1-x)
\label{dirgammaPDF}
\end{equation}
and
\begin{equation}
f_a^{\gamma,\mrm{VMD}}(x,\mu^2)
= \sum_{V = \rho^0,\omega,\phi,\Jpsi}
\frac{4\pi\alpha}{f_V^2} f_a^{V}(x,\mu^2) ~.
\label{VMDgammaPDF}
\end{equation}
The anomalous part, finally, is fully calculable perturbatively, given
the boundary condition that the distributions should vanish for
$\mu^2 = p_0^2$. In principle, everything is therefore given.
In practice, the vector-meson distributions are not known, and so
one is obliged to make further assumptions, such as
$\rho^0 \approx \pi^0 \approx (\pi^+ + \pi^-)/2$. Since the $\rho^0$
is rather short-lived, it is not impossible that it could be somewhat
different from a $\pi$, e.g. with fewer partons at small $x$.
By comparison, conventional distributions are defined for resolved
processes only:
\begin{equation}
f_a^{\gamma,\mrm{res}}(x,\mu^2)  =  f_a^{\gamma,\mrm{VMD}}(x,\mu^2) +
f_a^{\gamma,\mrm{anom}}(x,\mu^2;p_0^2) ~.
\label{resgammaPDF}
\end{equation}
The resolved distributions are then assumed to be given in a
completely free way, at some input scale, i.e. without any direct
relation with the vector-meson distributions.

\section{Cross Sections}

Total hadronic cross sections show a characteristic fall-off at
low energies and a slow rise at higher energies. This behaviour
can be parametrized by the form
\begin{equation}
\sigma_{\mrm{tot}}^{AB}(s) = X^{AB} s^{\epsilon} + Y^{AB} s^{-\eta}
\label{sigmatotAB}
\end{equation}
for $A + B \to X$. The powers $\epsilon$ and $\eta$
are universal, with fit values \cite{DL92}
\begin{equation}
  \epsilon \approx 0.0808 ~, \qquad
  \eta \approx 0.4525 ~,
\label{epsivalue}
\end{equation}
while the coefficients $X^{AB}$ and $Y^{AB}$ are
process-dependent.
Equation (\ref{sigmatotAB}) can be interpreted within Regge theory, where
the first term corresponds to pomeron exchange and
gives the asymptotic rise of the cross section. Ultimately,
this increase violates the Froissart--Martin bound \cite{Froissart};
$\epsilon$ should therefore be thought of as slowly decreasing with
energy (owing to multi-pomeron exchange effects), although data at
current energies are well fitted by a constant $\epsilon$.
The second term, the reggeon one, is mainly of interest at low
energies. For the purpose of our study we do not rely on the Regge
interpretation of eq.~(\ref{sigmatotAB}), but can merely consider it as
a convenient parametrization.

The VMD part of the $\gamma\p$ cross section should have a similar
behaviour, but there is no compelling reason why the direct and
anomalous parts would have to. However, empirically, the $\gamma\p$
data are well described by
\begin{equation}
\sigma_{\mrm{tot}}^{\gamma\p}(s) \approx 67.7 \, s^{\epsilon} +
129 \, s^{-\eta} ~~[\mu\mrm{b}],
\label{sigtotgap}
\end{equation}
with $s$ in GeV$^2$. (Cross-sections are throughout given in
mb for hadron--hadron interactions, in $\mu$b for $\gamma$--hadron
ones and in nb for $\gamma\gamma$ ones.) Actually, the above formula
is a prediction \cite{DL92} preceding the HERA data \cite{HERAtot}.

If we take the Regge-theory ansatz seriously also for the photon,
it is possible to derive an expression for the total $\gamma\gamma$
cross section
\begin{equation}
\sigma_{\mrm{tot}}^{\gamma\gamma}(s) \approx 211 \, s^{\epsilon} +
297 \, s^{-\eta} ~~[\mrm{nb}].
\label{sigtotgaga}
\end{equation}
This is based on the assumption
that the pomeron and reggeon terms factorize,
$X^{AB} = \beta_{A\pom} \beta_{B\pom}$ and
$Y^{AB} = \gamma_{A\reg} \gamma_{B\reg}$, so that
$X^{\gamma\gamma} = (X^{\gamma\p})^2/X^{\p\p}$ and
$Y^{\gamma\gamma} = (Y^{\gamma\p})^2/Y^{\p\p}$,
with $X^{\p\p} \approx 21.70$ and $Y^{\p\p} \approx 56.08$.
In hadronic cross sections, factorization seems valid for the pomeron
term but not for the reggeon one, e.g. $X^{\pbar\p} = X^{\p\p}$ while
$Y^{\pbar\p} \approx 98.39 \gg Y^{\p\p}$. An equally valid guess for
$Y^{\gamma\gamma}$ would then be obtained by $Y^{\gamma\gamma} =
2 (Y^{\gamma\p})^2/(Y^{\p\p} + Y^{\pbar\p}) \approx 215$. The
uncertainty in $Y^{\gamma\gamma}$ only affects the low-energy
behaviour, and so is not critical for us.

Note that factorization is assumed to hold separately for the pomeron
and the reggeon terms, not for the total cross section itself. That is,
the relation $\sigma_{\mrm{tot}}^{\gamma\gamma} =
(\sigma_{\mrm{tot}}^{\gamma\p})^2/\sigma_{\mrm{tot}}^{\p\p}$
is not exact in this approach, although numerically it is a very
good approximation.

Our eq.~(\ref{sigtotgaga}) above should be compared with
the time-honoured expression $\sigma^{\gamma\gamma} = 240 + 270/W$
\cite{Rosner}. This corresponds to a critical pomeron, $\epsilon = 0$,
as was commonly assumed in the early seventies, and an $\eta = 0.5$,
but it is otherwise in the same spirit as our formula. Also numerically
the two closely agree at not too large energies.

One should remember that our expression (\ref{sigtotgaga}) is here
`derived' based on a simple Regge-theory ansatz that has no real
validity for the photon. Next we will proceed to study the
contributions of the individual event classes. The constraints that come
from $\gamma\p$ physics data then directly feed into constraints on
the contribution from these classes and therefore on the total
$\gamma\gamma$ cross section. At the end of the day we will therefore
show that a cross section behaving roughly like
eq.~(\ref{sigtotgaga}) should be a good approximation. In doing so,
the properties of the event classes are also fixed, to a large extent.

Based on the subdivision into event classes, the total $\gamma\p$ cross
section may be written as
\begin{equation}
\sigma_{\mrm{tot}}^{\gamma\p} = \sigma_{\mrm{VMD}}^{\gamma\p} +
\sigma_{\mrm{dir}}^{\gamma\p} + \sigma_{\mrm{anom}}^{\gamma\p}
\label{sigdividegp}
\end{equation}
and the total $\gamma\gamma$ one as
\begin{equation}
\sigma_{\mrm{tot}}^{\gamma\gamma} =
\sigma_{\mrm{VMD}\times\mrm{VMD}}^{\gamma\gamma} +
2 \sigma_{\mrm{VMD}\times\mrm{dir}}^{\gamma\gamma} +
2 \sigma_{\mrm{VMD}\times\mrm{anom}}^{\gamma\gamma} +
\sigma_{\mrm{dir}\times\mrm{dir}}^{\gamma\gamma} +
2 \sigma_{\mrm{dir}\times\mrm{anom}}^{\gamma\gamma} +
\sigma_{\mrm{anom}\times\mrm{anom}}^{\gamma\gamma} ~.
\label{sigdividegg}
\end{equation}
Here we explicitly keep the factor of 2 for the off-diagonal terms,
where the r\^ole of the two incoming photons may be interchanged.

The $V\p$ cross sections may be parametrized as
\begin{eqnarray}
\sigma_{\mrm{tot}}^{\rho^0\p} \approx \sigma_{\mrm{tot}}^{\omega\p}
  & \approx & \frac{1}{2} \left( \sigma_{\mrm{tot}}^{\pi^+\p} +
  \sigma_{\mrm{tot}}^{\pi^-\p} \right)
  \approx 13.63 \, s^{\epsilon} + 31.79 \, s^{-\eta} ~~[\mrm{mb}],
\label{sigrho} \\
\sigma_{\mrm{tot}}^{\phi\p} & \approx &
  \sigma_{\mrm{tot}}^{\K^+\p} + \sigma_{\mrm{tot}}^{\K^-\p} -
  \sigma_{\mrm{tot}}^{\pi^-\p}
  \approx 10.01 \, s^{\epsilon} - 1.51 \, s^{-\eta} ~~[\mrm{mb}].
\label{sigphi}
\end{eqnarray}
(The $\Jpsi\p$ cross section is taken to be about a tenth of the
$\phi\p$ one, with a large amount of uncertainty; it is included in
the complete analysis but is neglected in our discussion here.)
Again using factorization for the pomeron and reggeon terms separately,
the total cross section for two vector mesons is
\begin{equation}
\sigma_{\mrm{tot}}^{V_1 V_2} \approx
\frac{X^{\p V_1} X^{\p V_2}}{X^{\p\p}} \, s^{\epsilon} +
\frac{Y^{\p V_1} Y^{\p V_2}}{Y^{\p\p}} \, s^{-\eta} ~.
\end{equation}

For a description of VMD events, a further subdivision into elastic
(el), diffractive (sd and dd for single and double diffractive)
and non-diffractive (nd) events is required. Keeping only the
simplest diffractive topologies, one may write
\begin{equation}
\sigma_{\mrm{tot}}^{AB}(s) = \sigma_{\mrm{el}}^{AB}(s) +
\sigma_{\mrm{sd}(XB)}^{AB}(s) + \sigma_{\mrm{sd}(AX)}^{AB}(s) +
\sigma_{\mrm{dd}}^{AB}(s) + \sigma_{\mrm{nd}}^{AB}(s)~.
\end{equation}
The elastic and diffractive cross sections for all required $V\p$ and
$V_1 V_2$ processes have been calculated and parametrized in the
context of our model presented in ref. \cite{haha}. The
non-diffractive cross-section is then given by whatever is left.
The $\sigma_{\mrm{nd}}$ may be further subdivided into a low-$\pT$
and a high-$\pT$ class. Since the $2\to2$ parton--parton scattering
cross sections are divergent in the limit $\pT \to 0$, some further
care is needed for this classification. We expect the perturbative
formulae to break down at small $\pT$, since an exchanged gluon
with a large transverse wavelength $\lambda_{\perp} \sim 1 / \pT$
cannot resolve the individual colour charges inside a hadron.
The hadron being a net colour singlet, the effective
coupling should therefore vanish in this limit.
A parameter $\pTmin$ is introduced to describe the border
down to which the perturbative expression is assumed to be valid
\cite{gammap}:
\begin{equation}
\pTmin(s) = \pTmin^{\mrm{VMD}}(s) \approx 1.3 + 0.15 \,
\frac{\ln(E_{\mrm{cm}}/200)}{\ln(900/200)} ~~[\mrm{GeV}]~.
\label{pTmins}
\end{equation}
The jet rate above $\pTmin$ may still be large, in fact
even larger than the total $\sigma_{\mrm{nd}}$. It is therefore
necessary to allow for the possibility of having several perturbative
parton--parton interactions in one and the same event, i.e. to
unitarize the jet emission probability. We do this
using the formalism of ref. \cite{TSMZ}.

\begin{figure}[tbp]
\captive%
{The total VMD$\times$VMD cross section, full curve, and its subdivision
by vector-meson combination. The components are separated by dashed
curves, from bottom to top: $\rho^0\rho^0$, $\rho^0\omega$, $\rho^0\phi$,
$\rho^0\Jpsi$, $\omega\omega$, $\omega\phi$, $\omega\Jpsi$, $\phi\phi$,
$\phi\Jpsi$, and $\Jpsi\Jpsi$. Some of the latter components are too
small to be resolved in the figure.
\label{Fig3}}\\[8mm]
\captive%
{The total VMD$\times$VMD cross section, full curve, and its subdivision
by event topology. The components are separated by dashed curves, from
bottom to top: elastic, single diffractive (split for the two sides by
the dotted curve), double diffractive, and non-diffractive (including
jet events unitarized).
\label{Fig4}}
\end{figure}

The total VMD cross sections are obtained as weighted sums of
the allowed vector-meson states,
\begin{eqnarray}
\sigma_{\mrm{VMD}}^{\gamma\p} & = &
\sum_V \frac{4\pi\alphaem}{f_V^2} \, \sigma_{\mrm{tot}}^{V\p}
\approx 53.4 \, s^{\epsilon} + 115 \, s^{-\eta} ~~[\mu\mrm{b}],
\label{sigVMDgap} \\
\sigma_{\mrm{VMD}\times\mrm{VMD}}^{\gamma\gamma} & = &
\sum_{V_1} \frac{4\pi\alphaem}{f_{V_1}^2}
\sum_{V_2} \frac{4\pi\alphaem}{f_{V_2}^2} \,
\sigma_{\mrm{tot}}^{V_1 V_2}
\approx 131 \, s^{\epsilon} + 236 \, s^{-\eta} ~~[\mrm{nb}].
\label{sigVMDgaga}
\end{eqnarray}
In Fig.~\ref{Fig3} we show the breakdown of
$\sigma_{\mrm{VMD}\times\mrm{VMD}}^{\gamma\gamma}$
by vector-meson combination. Obviously the $\rho^0\rho^0$
combination dominates. The same kind of formulae as above also
apply for the subdivision into elastic, diffractive and
non-diffractive events. This subdivision is shown in Fig.~\ref{Fig4}
for the sum of all meson combinations, which then mainly reflects
the $\rho^0\rho^0$ composition.

\begin{figure}[tbp]
\captive%
{The total $\gamma\p$ cross section and its subdivision
by event topology. Full curve: the parametrization of
eq.~(\protect\ref{sigtotgap}). The dashed curves, from
bottom to top: VMD, VMD+direct and VMD+direct+anomalous.
\label{Fig5}}
\end{figure}

Comparing eqs. (\ref{sigtotgap}) and (\ref{sigVMDgap}), about 80\%
of the $\gamma\p$ total cross section is seen to come from the VMD
term. The remaining 20\% is to be attributed to the direct and anomalous
components. At small energies the anomalous part is negligible, and so
the dependence of the direct cross section on $p_0$ can be used to
determine this parameter. We obtain a value of $p_0 \approx 0.5$~GeV,
which is consistent with the simple-minded answer
$p_0 \approx m_{\phi}/2$, and also gives a reasonable
$f_a^{\gamma,\mrm{res}}(x,\mu^2)$ \cite{gammap}.
The anomalous process contains two cut-off parameters, the $p_0$ scale
for the photon to branch to a perturbative $\q\qbar$ pair and a
$\pTmin^{\mrm{anom}}$ scale for one of the anomalous-photon partons
to interact in a hard process. As a first guess, one might choose
$\pTmin^{\mrm{anom}}$ also to be given by eq.~(\ref{pTmins}).
However, this turns out to give a cross section increasing too rapidly.

Physically, it is understandable why hard processes should be
more suppressed at small $\pT$ in anomalous processes than in VMD ones:
the anomalous photon corresponds to a $\q\qbar$ pair of larger virtuality
than a VMD one, and hence of smaller spatial extent. The best recipe for
including this physics aspect is not well understood.
Remembering that the anomalous cross section is the product
(or, more precisely, the convolution) of the anomalous parton distributions
and the hard partonic $2\rightarrow 2$ scattering cross sections, one can,
purely pragmatically, imagine two extreme procedures to weaken
the too-strong rise of $\sigma^{\gamma \p}_{\mrm{anom}}$: either
reduce the partonic cross section by increasing $\pTmin^{\mrm{anom}}$,
or decrease the values of the anomalous parton distributions
by choosing a smaller value for the scale $\mu$, compare
eq.~(\ref{gammaPDF}).
Over the HERA energy range, say 100~GeV $\leq E_{\mrm{cm}}
\leq$ 300~GeV, both choices
\begin{eqnarray}
\pTmin^{\mrm{anom}}(s) = & 1.5 + 0.0035 \, E_{\mrm{cm}} ~~[\mrm{GeV}]
\qquad & ; \qquad \mu = \pT
\label{pTminanom}\\
\pTmin^{\mrm{anom}}(s) = & \pTmin^{\mrm{VMD}}(s)
\qquad & ; \qquad \mu = \frac{\pT}{r} \quad , \quad r \approx
 \pTmin^{\mrm{VMD}}/p_0
\label{pTminanomnew}
\end{eqnarray}
give sensible answers, whereof we will use (\ref{pTminanom}) as our
main option. The resulting subdivision of the
$\gamma\p$ total cross section is shown in Fig.~\ref{Fig5}.

\begin{figure}[tbp]
\captive%
{Comparison of $\gamma\gamma$ partial cross sections obtained
by integration, full curves, and by a simple factorization
ansatz, dashed curves. In a few of the plots, dotted curves give
variations of the main curve obtained by integration, see text.
\label{Fig6}}
\end{figure}

Turning to the $\gamma\gamma$ cross sections, in principle all
free parameters have now been fixed, and the cross section for each of
the six event classes can be obtained. The VMD$\times$VMD one has
already been discussed; the others are given as integrals of $2 \to 2$
scattering cross sections above the respective $\pT$ cut-offs already
specified. The results are shown in Fig.~\ref{Fig6}, class by class. For
comparison, we also show the results that would be obtained if the
simple factorization ansatz used to derive
$\sigma_{\mrm{tot}}^{\gamma\gamma}$ in eq.~(\ref{sigtotgaga})
is valid component by component of the photon wave function. That is,
if the composition of the $\gamma\p$ cross section at some energy is
80\% VMD, 15\% direct and 5\% anomalous, say, the $\gamma\gamma$ cross
section is then assumed to be $0.8 \times 0.8 = 64$\% VMD$\times$VMD,
$2 \times 0.8 \times 0.15 = 24$\% VMD$\times$direct, etc. This relative
composition is then scaled by the assumed
$\sigma_{\mrm{tot}}^{\gamma\gamma}$ of eq.~(\ref{sigtotgaga})
to get actual cross section numbers.

A few comments about each of the classes:
\begin{Enumerate}
\item For the VMD$\times$VMD class
in principle the simple factorization ansatz is exact in our model.
The small deviations observed in Fig.~\ref{Fig6}a are not to be taken
seriously, they come from the fact that the numerically integrated
$\sigma_{\mrm{VMD}}^{\gamma\p} + \sigma_{\mrm{dir}}^{\gamma\p} +
\sigma_{\mrm{anom}}^{\gamma\p}$ does not agree perfectly with the
desired $\sigma_{\mrm{tot}}^{\gamma\p}$ of eq.~(\ref{sigtotgap}),
cf. Fig.~\ref{Fig5}.
Therefore the relative $\gamma\p$ composition used in the factorization
ansatz has been slightly rescaled so as to add to unity.
\item Agreement is also acceptable for VMD$\times$direct. This is not
so surprising, since this process is just a scaled-down version of
a direct process on a vector meson target. Disagreements should
therefore primarily come from differences between the meson and the
proton structure functions. (The agreement would be perfect
if the experimental input had come from $\rho^0\rho^0$ and
$\gamma\rho^0$, and so forth for the other mesons,
rather than from $\p\p$ and $\gamma\p$.) We show an example of the
variation that may come from using different sets of
parton-distribution functions for the pion. The similarity of the
large-energy behaviour reflects the fact that we have used the
same small-$x$ modification for both sets \cite{gammap}.
\item Also the VMD$\times$anomalous component agrees well; again this
class can be seen as a scaled-down version of the anomalous $\gamma\p$
class. The dependence on the choice of parton distributions is reduced,
compared with the previous class, since anomalous processes involve
larger $x$ and $\mu^2$ (i.e. better constrained regions) than the
direct ones.
\item The direct$\times$direct component is not at all well predicted
by the factorized ansatz. The latter yields a cross
section growing at large energies at a rate related to the
small-$x$ behaviour of the proton distribution functions, i.e.\
$\propto s^{\epsilon}$ for our modified distributions. On the other
hand, the total cross section for
$\gamma\gamma \to \q\qbar$ is proportional to $\ln(s/p_0^2)/s$,
and thus drops rapidly with c.m. energy. Charm production has here
been included using massive matrix elements (whereas masses
are neglected for a few other processes), since the
assumed $p_0$ cut-off is smaller than the charm mass and since charm
makes up a significant fraction of the total contribution.
\item The direct$\times$anomalous component again compares reasonably
well with the prediction from factorization. There is some ambiguity
about the choice of
$\pT$ cut-off for hard scatterings, but it is seen that results are
not drastically sensitive to this, even when the default
$\pTmin^{\mrm{anom}}$ cut-off of the hard process is replaced by $p_0$.
\item The anomalous$\times$anomalous process, finally, is most uncertain,
because its cross section sensitively depends
on the choice of method to keep $\sigma_{\mrm{anom}}^{\gamma\p}$ low.
Procedure (\ref{pTminanomnew}) yields a much smaller cross section
than (\ref{pTminanom}), which is easily understood by recalling that
$\f_a^{\gamma,\mrm{anom}}$ enters quadratically in the
anomalous$\times$anomalous cross section.
If method (\ref{pTminanomnew})
is the correct one then, in fact, factorization holds
to a very good approximation.
Also other mechanisms could be invoked to argue for a smaller
$\sigma_{\mrm{anom}\times\mrm{anom}}^{\gamma\gamma}$ than the default
one, such as the possibility of several parton--parton interactions
in the same event.
\end{Enumerate}

When taking the sum of the six classes above, eq.~(\ref{sigdividegg}),
it should be remembered that the first three are the
dominant ones. In fact, since the direct and anomalous components
together give about 20\% of the
$\gamma\p$ total cross section, the expectation is that the last
three classes together would only give a 4\% contribution to the
total $\gamma\gamma$ cross section. Apart from the large uncertainty
in the anomalous$\times$anomalous component, this is also the way it
works out. The first three classes, on the other hand, are all
related to the respective $\gamma\p$ classes, with only a replacement
of a $\p$ by a $V$ (and an extra weight factor $4\pi\alphaem/f_V^2$).
This makes the argumentation for eq.~(\ref{sigtotgaga}) credible,
in spite of the absense of a (well-defined) coupling between a direct
photon and a pomeron.

\begin{figure}[tbp]
\captive%
{The total $\gamma\gamma$ cross section. Full curve: the
parametrization of eq.~(\protect\ref{sigtotgaga}).
Dashed curve: result from sum of integrations of the six components.
Data points: open triangles PLUTO 1984, filled triangles PLUTO 1986,
squares TPC/2$\gamma$ 1985, spades TPC/2$\gamma$ 1991, circles MD-1
1991 \protect\cite{MD1}.
\label{Fig7}}
\end{figure}

The $\sigma_{\mrm{tot}}^{\gamma\gamma}$ obtained by integration
of the six components is compared with eq.~(\ref{sigtotgaga}) and
experimental data in Fig.~\ref{Fig7}. As already discussed, the results
of the integration are uncertain by some amount, so within this
band of uncertainty the agreement with eq.~(\ref{sigtotgaga})
is very good. It is also readily seen that data are not
(yet) precise enough to provide any real constraints, but are in
generally good agreement with both approaches.

One can also compare our $\sigma_{\mrm{tot}}^{\gamma\gamma}$
with the numbers obtained in various minijet-based
approaches \cite{minijet}. For $E_{\mrm{cm}} = 200$~GeV, cross sections
in the range 1000--1800 nb are typically obtained, but are reduced to
about 500 nb if unitarity is enforced, in agreement with our results.

\section{Event Properties}

The subdivision of the total $\gamma\p$ and $\gamma\gamma$ cross
sections above, with the related choices of cut-off parameters etc.,
specifies the event composition at the hard-scattering level. For
studies of the complete event structure, it is necessary to add models
for initial- and final-state QCD radiation (parton showers), for beam
remnants, and for fragmentation and secondary decays \cite{gammap}.
A Monte Carlo generation of complete hadronic final states is obtained
with {\sc Pythia}/{\sc Jetset} \cite{PyJe}.
Thus any experimental quantity can be studied. This section gives some
representative examples. In particular, we compare the properties
of $\p\p$, $\gamma\p$ and $\gamma\gamma$ events. It should be noted
that $\p\p$ and $\pbar\p$ events are very similar for the
quantities studied here. Unless otherwise specified, the figures
refer to an $E_{\mrm{cm}} = \sqrt{s_{\gamma\gamma}} = 100$~GeV. As we
will show at the end of the section, the qualitative features do not
depend critically on this choice.

\begin{figure}[tbp]
\captive%
{The total transverse energy per event, separately normalized for each of the
six event classes. Top frame: VMD$\times$VMD: full histogram;
VMD$\times$direct: dashed one; and VMD$\times$anomalous: dash-dotted
one. Bottom frame: direct$\times$direct: full histogram;
direct$\times$anomalous: dashed one; and anomalous$\times$anomalous:
dash-dotted one.
\label{Fig8}}
\end{figure}

Figure~\ref{Fig8} shows the $\sum E_{\perp}$ per event for each
of the six components of the $\gamma\gamma$ cross section. The spike
at small $\sum E_{\perp}$ for the VMD$\times$VMD class comes from
elastic scattering events, e.g. $\gamma\gamma \to \rho^0\rho^0$.
Also diffractive events contribute in this region. The
large-$\sum E_{\perp}$ tail of the VMD$\times$VMD curve is enhanced
by the possibility of multiple parton--parton interactions, which is
only included for this class. Because of the
larger $\pTmin^{\mrm{anom}}$ cut-off, the classes involving anomalous
photons typically have larger $\sum E_{\perp}$, while the smaller
$p_0$ cut-off for the direct processes corresponds
to smaller median $\sum E_{\perp}$. However, note that
the $\gamma\gamma \to \q\qbar$ processes only fall off very slowly with
$\pT$, in part because of the absense of structure functions, in part
because of the form of the matrix element itself. The
direct$\times$direct class therefore wins out at very large
$\sum E_{\perp}$.

\begin{figure}[tbp]
\captive%
{The total transverse energy per event for different beams:
$\gamma\gamma$: full histogram; $\gamma\p$: dashed one: and
$\p\p$: dash-dotted one.
\label{Fig9}}\\[8mm]
\captive%
{Transverse energy flow as a function of rapidity for different beams:
$\gamma\gamma$: full histogram; $\gamma\p$: dashed one; and
$\p\p$: dash-dotted one.
\label{Fig10}}
\end{figure}

The results of Fig.~\ref{Fig8} are a bit misleading, since the relative
importance of the six event classes is not visible. The weighted
mixture is shown in Fig.~\ref{Fig9}, also compared with $\gamma\p$ and
$\p\p$ events. One observes a steady progression, with
$\langle \sum E_{\perp} \rangle_{\p\p} <
\langle \sum E_{\perp} \rangle_{\gamma\p} <
\langle \sum E_{\perp} \rangle_{\gamma\gamma}$.
This pattern, of more activity for a $\gamma$ than for a $\p$,
is seen in essentially all distributions. The elastic spike at small
$\sum E_{\perp}$ is less pronounced for $\gamma\gamma$, due to three
factors: the VMD$\times$VMD component is only a part of the
$\gamma\gamma$ cross
section, elastic scattering is a smaller fraction of the total
$\rho^0\rho^0$ cross section than it is for $\p\p$, and kinetic
energy in the $\rho^0 \to \pi^+\pi^-$ decays add to the total
transverse energy.

The $E_{\perp}$ flow as a function of rapidity, $\d E_{\perp} / \d y$,
is given in Fig.~\ref{Fig10}.
It illustrates how $\gamma\p$ interpolates between $\p\p$ and
$\gamma\gamma$: around the direction of the incoming photon, the
$\gamma\p$ events look like the $\gamma\gamma$ ones, while they look
more like $\p\p$ ones in the opposite direction, with an intermediate
behaviour in the central region.

The charged-multiplicity distributions follow essentially the same
pattern as shown for the $\sum E_{\perp}$ ones in Figs.~\ref{Fig8},
\ref{Fig9} and \ref{Fig10},
and are therefore not included here. There is one noteworthy
exception, however: the direct$\times$direct component does not have
a tail out to large multiplicities. That is, even if the process
$\gamma\gamma \to \q\qbar$ can generate large $\pT$ values, the
absence of any beam jets keeps the multiplicity down.

\begin{figure}[tbp]
\captive%
{Charged particle inclusive $\pT$ spectra for different beams:
$\gamma\gamma$: full histogram; $\gamma\p$: dashed one; and
$\p\p$: dash-dotted one.
\label{Fig11}}\\[8mm]
\captive%
{Jet rate as function of the transverse jet energy for different
beams: $\gamma\gamma$: full histogram; $\gamma\p$: dashed one; and
$\p\p$: dash-dotted one.
\label{Fig12}}
\end{figure}

The transverse momentum spectrum of charged particles is shown in
Fig.~\ref{Fig11}. The larger high-$\pT$ tail of the $\gamma\gamma$
processes is one of the simplest observables to experimentally
establish differences between $\p\p$, $\gamma\p$ and $\gamma\gamma$.
Of course, the cause of the differences is to be sought in the higher
jet rates associated with photon interactions. The jet spectra are
compared in Fig.~\ref{Fig12}, using a simple cone algorithm where
a minimum $E_{\perp}$ of 5 GeV is required inside a cone of
$\Delta R = \sqrt{ (\Delta\eta)^2 + (\Delta\phi)^2} < 1$.
Already for an $E_{\perp\mrm{jet}}$ of 5 GeV there are about ten times
as many jets in  $\gamma\gamma$ as in $\p\p$, and this ratio then
increases with increasing $E_{\perp\mrm{jet}}$.

\begin{figure}[tbp]
\captive%
{Transverse energy flow for $E_{\mrm{cm}} = 25$~GeV as a
function of rapidity for different beams:
$\gamma\gamma$: full histogram; $\gamma\p$: dashed one; and
$\p\p$: dash-dotted one.
\label{Fig13}}\\[8mm]
\captive%
{Transverse energy flow for $E_{\mrm{cm}} = 400$~GeV as a
function of rapidity for different beams:
$\gamma\gamma$: full histogram; $\gamma\p$: dashed one; and
$\p\p$: dash-dotted one.
\label{Fig14}}
\end{figure}

To illustrate the energy dependence of these distributions,
Figs.~\ref{Fig13} and \ref{Fig14} give the
$\d E_{\perp} / \d y$ flow for $\gamma\gamma$ c.m. energies
of 25 and 400 GeV, respectively. These can be compared with the
result for 100 GeV in Fig.~\ref{Fig10}. Qualitatively, the same pattern
is seen at all three energies, although relative differences tend to be
somewhat reduced at larger energies. This is also true for other
observables, such as jet rates. One reason is that the possibility of
multiple parton--parton interactions in the VMD component pushes
up the activity in those events at larger energies, and thus brings
them closer to the anomalous class. The importance of the direct class,
on the other hand, is reduced at large energies. Further, at large
energies, jet production is dominantly initiated by small-$x$
incoming partons, where the VMD and anomalous distributions are
more similar than at large $x$ (although still different).

\section{Summary}

In this paper we have shown that our model for $\gamma\p$ events
\cite{gammap} can be consistently generalized to $\gamma\gamma$ events.
That is, essentially all free parameters are fixed by
(low-energy) $\gamma\p$
phenomenology. Since we start out with a more detailed subdivision
of the $\gamma\p$ total cross section than has conventionally been
done in the past, our $\gamma\gamma$ model also contains a
richer spectrum of possible processes. We distinguish six main event
classes, but most of these contain further subdivisions. The aim is
that this approach will allow predictions for a broader range of
observables than is addressed in conventional models.
For instance, although not discussed in detail here, our approach
does correlate the hard-jet physics in the central rapidity region
with the structure of the beam remnants.

This does not mean that all results are complicated. We have shown
that the simple Regge-theory expression
$\sigma_{\mrm{tot}}^{\gamma\gamma}(s) \approx
211 \, s^{0.08} + 297 \, s^{-0.45}$~[nb] comes very close to what is
obtained in our full analysis. We therefore expect this expression
to be good to better than 10\% from a few GeV onwards, at least to
the top $\gamma\gamma$ energies that could be addressed with the next
generation of linear $\ee$ colliders. Also global event properties
show a very simple pattern, with more activity (transverse energy,
multiplicity, jets, \ldots) in $\gamma\p$ events than in $\p\p$ ones,
and still more in $\gamma\gamma$ ones. This is perhaps contrary to
the na\"{\i}ve image of a `clean' point-like photon.
The $\gamma\p$ events show
their intermediate status by having a photon (proton) hemisphere that
looks much like $\gamma\gamma$ ($\p\p$) events, with a smooth
interpolation in the middle.

This does not mean that all problems have been solved. In particular,
the nature of the anomalous component of the photon is still not
well understood. This is reflected in the different cut-off procedures
that could be applied at small transverse momenta. Further, all the
discussions so far have been on incoming real photons; the transition
to the deeply-inelastic-scattering region has not been addressed so far.
More problems may well crop up once our model is compared with
observations. Therefore further data from HERA, TRISTAN, LEP 1 and 2,
and future linear $\ee$ colliders will have much to teach us.

\end{document}